# Tunable magnetic anisotropy, Curie temperature and band alignment of two-dimensional ferromagnet VSiSnN$_4$ via non-volatile ferroelectrical control

Kang-Jie Li[2], Ze-Quan Wang[1], Zu-Xin Chen[2,*], Yusheng Hou[1,*]

**AFFILIATIONS**

[1] Guangdong Provincial Key Laboratory of Magnetoelectric Physics and Devices, Center for Neutron Science and Technology, School of Physics, Sun Yat-Sen University, Guangzhou, 510275, China

[2] Guangdong Provincial Key Laboratory of Chip and Integration Technology School of Semiconductor Science and Technology, South China Normal University, Guangzhou 528225, China

## ABSTRACT

The emergence of multiferroic materials, which possess both ferromagnetic (FM) and ferroelectric (FE) properties, drive advancements in magnetoelectric applications and the next generation of spintronics. Based on first-principles calculations, we investigate an engineered two-dimensional multiferroic van der Waals heterostructures consisting of FM VSiSnN$_4$ monolayer (ML) and fully hydrogenated FE AlN bilayer. We find that the magnetic anisotropy of VSiSnN$_4$ ML is tunable between out-of-plane and in-plane and a phase transition between semiconductor and metal is induced in VSiSnN$_4$/AlN bilayer when the FE polarization direction of AlN bilayer is reversed. Surprisingly, when the FE polarization of AlN bilayer is upward, the Curie temperature of VSiSnN$_4$/AlN bilayer can be significantly increased from 204K to 284K. Such non-volatile and tunable magnetic anisotropy, Curie temperature and band alignment in VSiSnN$_4$/AlN multiferroic heterostructure are highly promising for future low-current operation of data storage and logic devices.

Authors to whom correspondence should be addressed: houysh@mail.sysu.edu.cn and chenzuxin@m.scnu.edu.cn

Multiferroic materials[1-5] are characterized by the simultaneous presence of two or more ferroic orderings, such as (anti)ferromagnetic, (anti)ferroelectric, or ferroelastic behaviors. These materials enable the direct manipulation of magnetism (electric polarization) via electric field (magnetic field)[6-8], thereby revolutionizing the storage technology for electro-writing and magnetic-reading[9, 10]. This advancement leads to enhanced storage density and significantly reduced energy consumption. Consequently, constructing novel multiferroics and optimizing magnetoelectric coupling have been crucial research with profound implications for scientific progress and technological innovation [7, 11-13]. In terms of compositions, multiferroic materials can be categorized into two groups[10]: single-phase multiferroics[4, 14, 15] and artificial multiferroic heterostructures which are built from ferroelectric (FE) and magnetic materials[16-21]. Among these, the representative of single-phase multiferroics are type-I multiferroics $BiFeO_3$[22-26] and type-II multiferroics $TbMnO_3$[27-31]. In type-I multiferroic materials, the magnetoelectric coupling is weak due to the distinct origins of ferromagnetism and ferroelectricity [22, 32]. On the other hand, in type-II multiferroics, the ferroelectricity originates from a super spin current formed by specific spin-ordered states along with spatial polarization resulting from charge order[33, 34]. However, the scarcity of single-phase materials with high magnetic transition temperatures and strong magnetoelectric coupling has led to the primary focus on artificial heterostructures composed of FE and FM materials[15, 35].

Recently, the emergence of van der Waals (vdW) FM and FE materials, such as $CrI_3$[36], $Fe_3GeTe_2$[37], $CrSBr$[38], $CuInP_2S_6$[39, 40] and $In_2Se_3$[41], open up new possibilities for the development of multiferroic heterostructures. These materials can maintain their FM or FE properties even in the monolayer (ML) limit. Furthermore, based on the successful fabrication of multilayer $MSi_2N_4$ ($M$=Mo, W)[42, 43], other two-dimensional (2D) MLs similar to $MSi_2N_4$ have been predicted using density functional theory (DFT) calculations[44, 45]. Notably, V$SiX$N$_4$ ($X$=Si, Sn) MLs have in-plane magnetic anisotropy and are narrow band gap semiconductors with a Curie temperature of more than 200 K [46], which significantly exceeds the Curie temperatures of $CrI_3$ ML and $Fe_3GeTe_2$ bilayer.

However, the characteristics of in-plane magnetization and low temperature of VSi$X$N$_4$ limit their application in spintronics. Various ways have been proved to be able to tune magnetic anisotropy and other properties, such as bending materials[47] and constructing heterostructure[48]. Among these methods, constructing multiferroic heterostructures can achieve both magnetic regulation and non-volatile purposes. Besides, it has been demonstrated that ML and double layers of wurtzite phase AlN flakes[49, 50] exhibit remarkable stability and possess a robust intrinsic out-of-plane FE polarization. Remarkably, this polarization can be effectively switched under electric fields[51]. Interestingly, the lattice constants of VSiSnN$_4$ and fully hydrogenated FE AlN bilayer (Fig. 1d) are 3.06 Å and 3.08 Å respectively, which suggests that the influence of lattice mismatch can be disregarded in their heterostructures. Overall, these highlight the potential of VSiSnN$_4$/AlN multiferroic heterostructures as promising candidates for investigating strong magnetoelectric coupling effects.

In this study, we perform a systematic study of the electronic and magnetic properties of 2D vdW multiferroic heterostructures composed of VSiSnN$_4$ ML and AlN bilayer, using DFT calculations. We find that the magnetism of VSiSnN$_4$ ML is sensitive to the polarization direction of AlN bilayer. Three findings are achieved in our work: i) The magnetic anisotropy of VSiSnN$_4$ ML can be controlled by changing the polarization direction of AlN bilayer; ii) A phase transition from semiconductor to metal occurs in VSiSnN$_4$/AlN; iii) The Curie temperature VSiSnN$_4$ can be remarkably enhanced by constructing VSiSnN$_4$/AlN heterostructure. Our work not only verifies the feasibility of artificially fabricating composite multiferroic materials but also offers alternative material candidates for designing novel magnetic storage devices.

VSi$X$N$_4$ ML consists of seven atomic layers, with a VN$_2$ layer sandwiched between a Si-N and an X-N layer (Fig. 1a-1b). The space group of VSi$_2$N$_4$ is $P\bar{6}m2$, whereas that of VSiSnN$_4$ is $P3m1$ due to the breaking of the spatial inversion symmetry. The optimized lattice constants of VSi$_2$N$_4$ and VSiSnN$_4$ are 2.89 Å and 3.06 Å, respectively. Considering that the phonon spectrum has been previously calculated[52, 53], it is expected that both VSi$_2$N$_4$ and VSiSnN$_4$ exhibit structural stability. We first study the electronic and magnetic characteristics of the FM pristine VSiXN$_4$ ML. The spin-polarized band structures without spin-orbit coupling (SOC) (Fig. 1f-1g) indicate that VSi$_2$N$_4$ shows a direct bandgap feature, while VSiSnN$_4$ demonstrates an indirect bandgap behavior. Additionally, the energy difference between the K point and the K' point are marked with blue dotted lines. It is clear that the energy difference is zero and valley degeneracy exists in this case.

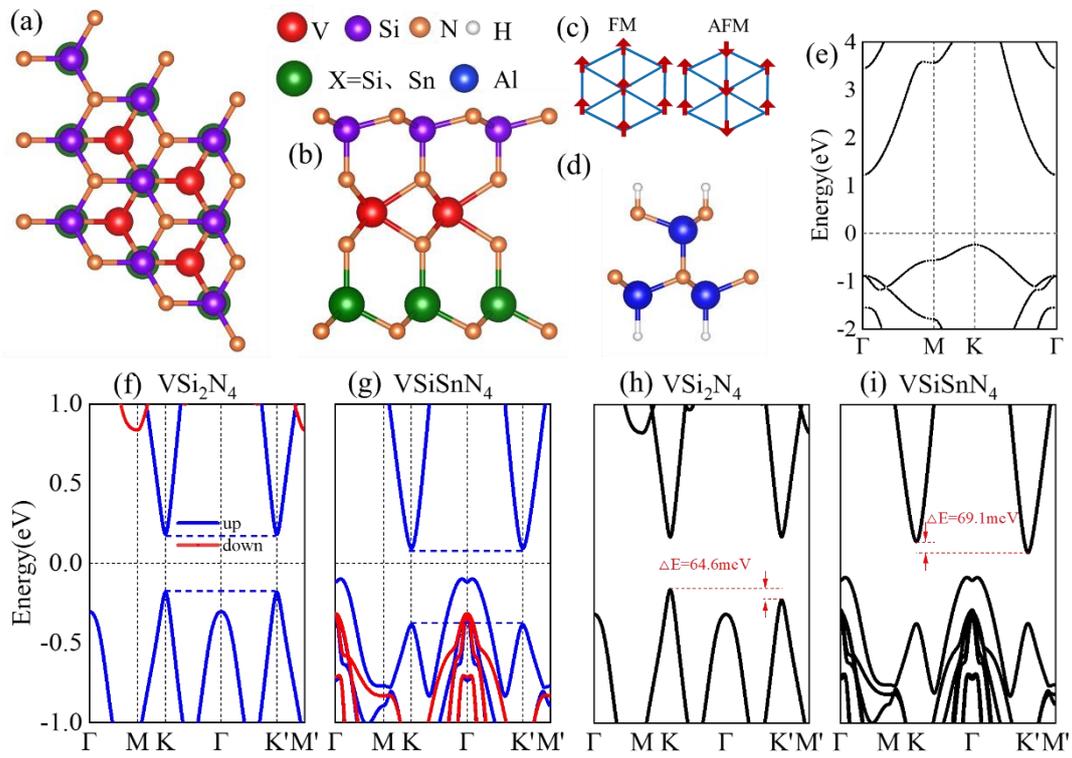

FIG. 1. Electric and magnetic properties of VSiXN$_4$ (X = Si, Sn) MLs and AlN bilayer. (a-b) Top and side views of the crystal structure of VSiXN$_4$. (c) Magnetic ground states of the ferromagnetic and antiferromagnetic. (d-e) Crystal structure and band structure of AlN bilayer. (f-g) Spin-polarized band structure of VSiXN$_4$. (h-i) Band structures of VSiXN$_4$ with SOC.

Fig. 1h-1i show the band structures of *VSiXN$_4$ (X = Si, Sn)* when SOC is considered. The energy difference between the valence band valleys of VSi$_2$N$_4$ and the conduction band valleys of VSiSnN$_4$ are 64.6 meV and 69.1 meV, which are represented by red dotted lines. Therefore, the introduction of SOC leads to valley splitting. The values of these splittings are listed in TABLE I. Besides, the analysis of the projection band reveals that valence and conduction bands in VSi$X$N$_4$ ML are predominantly from the $d$ orbital of V. The introduction of Sn in VSiSnN$_4$ weakens the contribution of $d_{z^2}$ orbital of V atom to the bottom of conduction band, compared with that of VSiSnN$_4$. Consequently, $d_{xy}$ and $d_{x^2-y^2}$ orbitals dominate the conduction band. Conversely, in the valence band, there is a significant reduction in contribution from $d_{xy}$ and $d_{x^2-y^2}$ orbitals, resulting in valence band primarily derived from V $d_{z^2}$ orbitals (Fig. S1).

To explore the magnetic properties of VSi$X$N$_4$, we employ a typical triangular lattice spin Hamiltonian consisting of Heisenberg exchange interactions and magnetic anisotropy energy (MAE). The spin Hamiltonian is as follows[54]:

$$H = J_1 \sum_{<ij>} \mathbf{S}_i \cdot \mathbf{S}_j + J_2 \sum_{\ll ij \gg} \mathbf{S}_i \cdot \mathbf{S}_j + J_3 \sum_{\lll ij \ggg} \mathbf{S}_i \cdot \mathbf{S}_j - K \sum_i (S_i^z)^2 \qquad (1),$$

where $S_i, S_j$ are the normalized spins at site $i, j$; $J_1, J_2, J_3$ present the nearest neighbor (NN), second-NN, and third-NN Heisenberg spin exchange parameters, respectively; $K$ is the magnetic anisotropy energy constant. Negative $J_n$ ($n = 1, 2, 3$) indicate FM interactions and a positive $K$ suggests an out-of-plane magnetic easy axis. DFT calculated $J_n$ and $K$ are summarized in TABLE I. As shown in TABLE I, the NN $J_1$ is FM and overwhelmingly dominant over $J_2$ and $J_3$ VSi$_2$N$_4$ and VSiSnN$_4$.

A comprehensive analysis of magnetic anisotropy energy ($E_{MAE}$) of magnets needs consider both magnetic shape anisotropy (MSA) energy ($E_{MSA}$) and magnetocrystalline anisotropy (MCA) energy ($E_{MCA}$), i.e., $E_{MAE} = E_{MCA} + E_{MSA}$. The emergence of MCA is attributed to SOC, while MSA originates from the dipole-dipole interaction. Generally, MAE in magnetic materials is primarily governed by MCA. However, in cases where there is a weak SOC, it becomes crucial to consider the contribution of MSA to MAE. This consideration arises due to the fact that MSA can surpass out-of-plane MCA under such circumstances, leading to a transition from an out-of-plane magnetic easy axis to in-plane one. $E_{MCA}$ is usually determined by the energy difference between in-plane and out-of-plane magnetizations in the presence of SOC, i.e., $E_{MCA} = E_{001} - E_{100}$, while $E_{MSA}$ can be calculated using the following formula[55]:

$$E_{MSA} = E_{001}^{Dipole-Dipole} - E_{100}^{Dipole-Dipole}$$

$$E^{Dipole-Dipole} = \frac{1}{2}\frac{\mu_0}{4\pi}\sum_{i \neq j}^{N} \frac{1}{r_{ij}^3} \times \left[\vec{M}_i \cdot \vec{M}_j - \frac{3}{r_{ij}^2}(\vec{M}_i \cdot \vec{r}_{ij})(\vec{M}_j \cdot \vec{r}_{ij})\right] \quad (2),$$

where $\vec{M}_i$ is the magnetic moments of V atoms and $\vec{r}_{ij}$ is a vector connecting two different V atoms at sites $i$ and $j$. As listed in TABLE I, the MAE of VSiSnN$_4$ ML exhibits an in-plane magnetic easy axis due to its stronger in-plane MSA ($E_{MSA} = 0.015 meV$) than its out-of-plane MCA ($E_{MCA} = -0.009 meV$). The in-plane magnetic anisotropy and the local magnetic moment (1.2 $\mu_B$) of V atoms suggest that VSi$X$N$_4$ ML exhibits characteristic properties of 2D XY magnets. The magnetic transition temperature can be obtained with $T_{BKT} = 1.335 J/k_B$, which are determined to be 410K and 208K for VSi$_2$N$_4$ and VSiSnN$_4$, respectively[56]. Here, only the NN $J_1$ is used to estimate $T_{BKT}$ because $J_2$ and $J_3$ are much weaker than $J_1$ (TABLE. S1) [52, 53].

TABLE I. Summary of magnetic and electric properties of VSi$X$N$_4$ and VSiSnN$_4$/AlN bilayer, including MCA (meV), MSA (meV), MAE (meV), $J_n$ (n=1, 2, 3) (meV), Gap (meV), $E_c^K - E_c^{K\prime}$ (meV) and T$_C$ (K).

| Structure | MCA | MSA | MAE | $J_1$ | $J_2$ | $J_3$ | Gap | T$_C$ | $E_c^K - E_c^{K\prime}$ |
|---|---|---|---|---|---|---|---|---|---|
| **VSi$_2$N$_4$** | 0.053 | 0.017 | 0.068 | -26.42 | -0.1 | -1.41 | 363 | 410.77 | 0 |
| **VSiSnN$_4$** | -0.009 | 0.015 | 0.008 | -13.55 | 0.1 | -0.25 | 190 | 208.22 | 69.1 |
| **HS-P↑ (Sn-terminal)** | -0.052 | 0.0187 | -0.0333 | -20.11 | -1.72 | 0.9 | 0 | 283.9 | 86.99 |
| **HS-P↓ (Sn-terminal)** | -0.015 | 0.015 | ~0 | -13.28 | 0.19 | -0.23 | 78 | 202.8 | 73.1 |
| **HS-P↑ (Si-terminal)** | -0.052 | 0.0182 | -0.0338 | -19.44 | -1.49 | 0.66 | 0 | 265.9 | -87.6 |
| **HS-P↓ (Si-terminal)** | -0.01 | 0.015 | 0.005 | -13.35 | 0.11 | -0.26 | 68 | 204.89 | -68.8 |

When constructing vertical VSiSnN$_4$/AlN heterostructures, we initially considered twelve stacking modes. By comparing their energies, four configurations are selected for further studies, and they were designated as HS(Sn-terminal)-P↑, HS(Sn-terminal)-

P↓, HS(Si-terminal)-P↑ and HS(Si-terminal)-P↓ (Fig. S2). The binding energy[57] ($E_b$) of these four heterostructures are verified using $E_b = E_{HS} - E_{VSiSnN_4} - E_{AlN}$, where $E_{HS}$, $E_{VSiSnN_4}$ and $E_{AlN}$ are total energies of heterostructure (HS), VSiSnN$_4$ and AlN bilayer, respectively. The calculated $E_b$ (TABLE. S1) suggests the potential for successful experimental synthesis of this vdW heterostructures. As for their stacking details, these four heterostructures can be classified into AA (upward FE polarization) and AB (downward FE polarization). The former is oriented along the Z direction by the *X* and Al atoms, while the latter is aligned by the H atoms with respect to the *X* atoms.

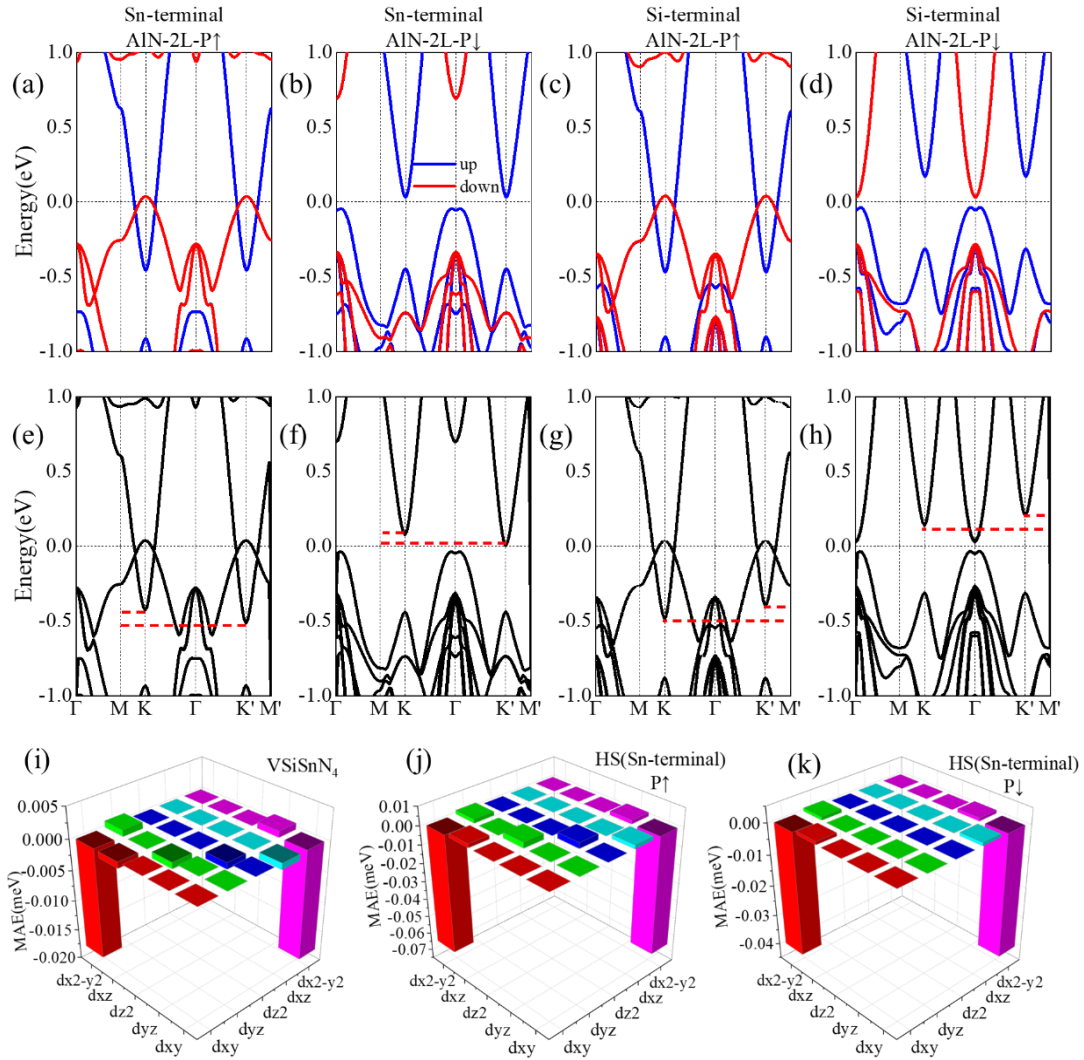

FIG. 2. (a-d) Spin-polarized band structure and (e-h) band structure with SOC for HS(Sn-terminal)-P↑, HS (Sn-terminal)-P↓, HS(Si-terminal)-P↑, HS(Si-terminal)-P↓,

respectively. (i-k) Orbital resolved change in magnetocrystalline anisotropy of V atom in VSiSnN$_4$ ML, HS(Sn-terminal)-P↑ and HS(Si-terminal)-P↑, respectively.

We first calculated the magnetic parameters of these four heterostructures, and results are shown in TABLE I. First, the NN $J_1$ is still FM and dominant over $J_2$ and $J_3$ in these four heterostructures. Compared with pristine VSiSnN$_4$ ML, an upward FE polarization enhances the NN $J_1$ by approximately 50% in HS(Sn-terminal)-P↑ and HS(Si-terminal)-P↑. Furthermore, significant enhancements are observed for both $J_2$ and $J_3$, specifically. These enhancements lead to a Curie temperature of 283.9K in HS(Sn-terminal)-P↑ and 265.9K in HS(Si-terminal)-P↑. When VSiSnN$_4$ is attached to AlN bilayer with a downward FE polarization, there is small alteration in the exchange coupling parameters and Curie temperatures.

We investigate the influence of FE polarization of AlN bilayer on the MAE of VSiSnN$_4$ ML. The MAE results are summarized in TABLE I. Remarkably, the upward FE polarization of AlN bilayer enhances both the MCA and MSA of VSiSnN$_4$ in heterostructures. Consequently, the magnetic easy axis is out-of-plane in VSiSnN$_4$/AlN, different from the pristine VSiSnN$_4$ ML. Conversely, when there is a downward FE polarization in AlN-2L, small changes are observed in MSA and MCA. Therefore, the magnetic easy axis of VSiSnN$_4$ remains within the plane in VSiSnN$_4$/AlN. These suggest that the incorporation of 2D magnetic materials into heterostructures with appropriate polarity layers can effectively modulate their MCA and MSA. To gain a deeper insight into the underlying mechanism of the tuned MAE of VSiSnN$_4$, we performed a comprehensive analysis on MSA and the Orbital-resolved MCAs by taking Sn-terminal heterostructures as an example. According to Eq. (2), MSA in this study is primarily influenced by parameter $\boldsymbol{M}$. As the lattice constant of VSiSnN$_4$ is fixed, $\boldsymbol{r}$ between magnetic V atoms is almost constant. When the FE polarization is upwards, there is an increase in the magnetic moment of V from 1.22 to 1.34 $\mu_B$, thereby resulting in an enhancement of MSA. It should be noted that under a downward polarization, the magnetic moment of V remains unchanged. For MCAs, we employ the second-order perturbation theory proposed by Wang et al.[58] to investigate the influence of hybridization between different orbitals of V on MCA (see details for obtaining orbital resolved MCA in SM), and corresponding results are depicted in Fig. 2i-2k. The determination of both sign and value of MCAs ultimately relies on the contributions from V 3$d$ orbitals with hybridization between $d_{xy}$ and $d_{x^2-y^2}$. When VSiSnN$_4$ ML is integrated with AlN bilayer, the hybridization of V 3$d$ orbitals increases, and the value of MCA rises following the sequence of HS(Sn-terminal)-P↑ > HS(Sn-terminal)-P↓ > VSiSnN$_4$ MLs.

Given that VSiSnN$_4$ is also a semiconductor, we have investigated the electronic properties of four heterostructures exhibiting a FM ground state. It is evident that, in comparison with the pristine VSiSnN$_4$ ML and AlN bilayer, both spin up and spin down channels cross the Fermi level, resulting in the metallic nature of HS(Sn-terminal)-P↑ and HS(Si-terminal)-P↑ (Fig. 2a-2c). Upon the polarization of AlN bilayer is downward, HS(Sn-terminal)-P↓ and HS(Si-terminal)-P↓ exhibit semiconductor behavior with bandgaps of 78 meV and 68 meV, respectively (Fig. 2b-2d). Such metal-semiconductor phase transition lies in the fact that both conduction band and valence band of HS (Sn-terminal)-P↓ originate from the spin-up channel, whereas the conduction band of HS(Si-terminal)-P↓ arises from the spin-down channel while its valence band originates from the spin-up channel. Therefore, by artificially manipulating the FE polarization direction of AlN bilayer, a metal-semiconductor phase transition can be induced in VSiSnN$_4$/AlN heterostructures.

Besides, we study the weighted band structure, which is shown in Fig. 3. One can see that when AlN bilayer has an upward FE polarization (Fig. 3a and 3e), its valence band and the conduction band of VSiSnN$_4$ intersect with the Fermi level, exhibiting a semi-metallic state. When the FE polarization of AlN bilayer is downward (Fig. 3c), the valence band and conduction band of HS (Sn-terminal)-P↓ originate from VSiSnN$_4$ ML, indicating the formation of a type-I band alignment (Fig. 3d). Nevertheless, the conduction band of HS (Si-terminal)-P↓ originates from AlN bilayer layer while its valence band is contributed by VSiSnN$_4$, thus belonging to a type-II band alignment (Fig. 3h).

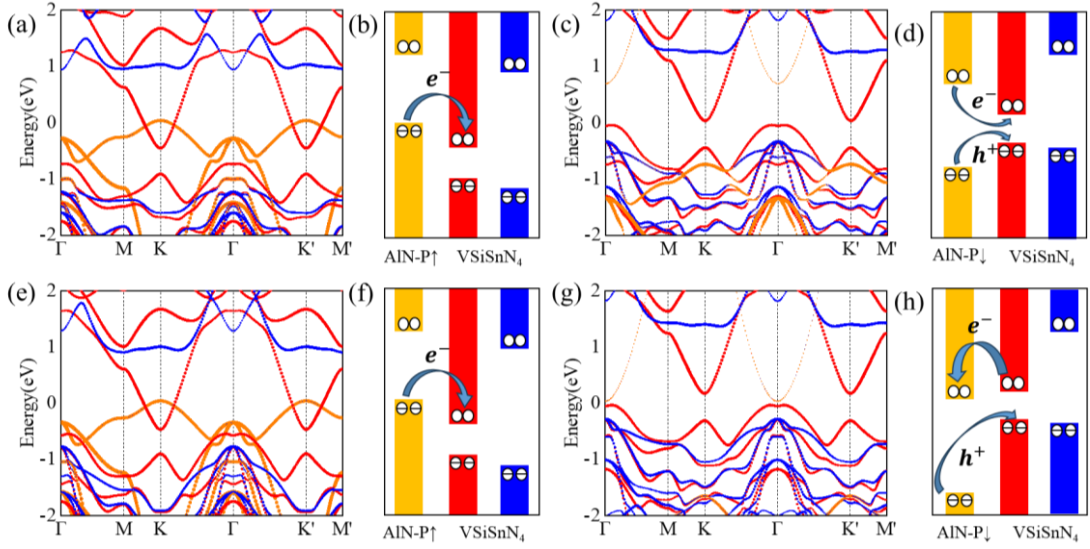

FIG. 3. Weighted band structures and band alignment diagram of (a, b) HS(Sn-terminal)-P↑, (c, d) HS (Sn-terminal)-P↓, (e, f) HS(Si-terminal)-P↑, (g, h) HS (Si-terminal)-P↓. Yellow lines represent bands of AlN, and red and blue lines represent spin-up and spin-down bands of VSiSnN$_4$, respectively.

Furthermore, we examine the electronic structures of the four heterostructures when SOC is considered (Fig. 2e-2h). We can see that the semiconductor-metal phase transitions and conduction valley splitting are maintained. The intriguing aspect is that the size and symbol of the valley split will vary depending on the polarization direction and contact interface (TABLE I). The energy difference of the conduction valley with out-of-plane magnetic anisotropy is expressed as[52]:

$$E_c^K - E_c^{K\prime} = i\langle d_{xy}|\lambda \hat{S}_z \hat{L}_z|d_{x^2-y^2}\rangle - i\langle d_{x^2-y^2}|\lambda \hat{S}_z \hat{L}_z|d_{xy}\rangle \qquad (3),$$

where $\hat{S}_z$ and $\hat{L}_z$ are spin angular and orbital angular operators, respectively. In the heterostructures with a Sn-terminal contact interface, it is observed that the magnetic moment of V atom is oriented vertically upwards, indicating a positive sign for $\hat{S}_z$ and validating $E_c^K - E_c^{K\prime} > 0$. Conversely, when constructing heterostructures at the Si-terminal, the magnetic moment of V atom aligns parallel to the -Z direction (Fig. S3), resulting in a negative value for $\hat{S}_z$ and $E_c^K - E_c^{K'}$. The findings imply that VSiSnN$_4$ could potentially find applications in the field of valley electronics[59].

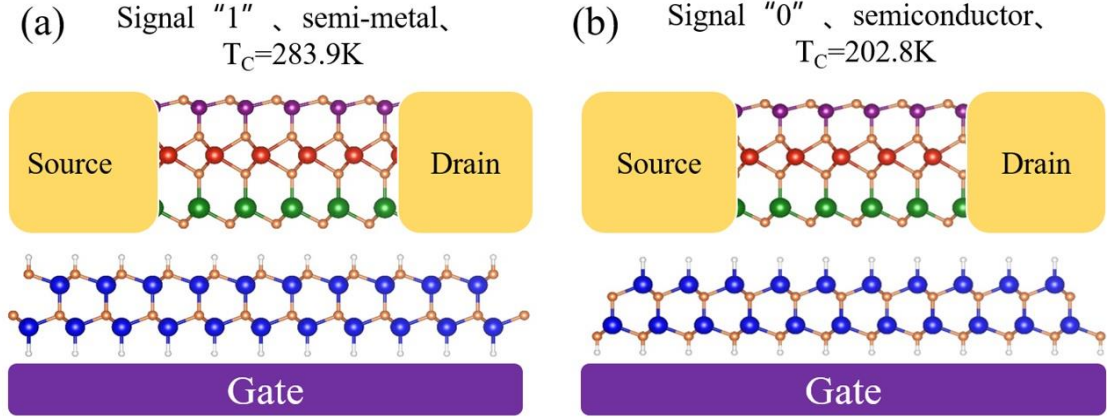

FIG. 4. (a, b) Non-volatile logic memory device diagram based on VSiSnN$_4$/AlN heterostructure.

Our study of VSiSnN$_4$/AlN heterostructures demonstrate the effective control of magnetic anisotropy, Curie temperature and phase transition in VSiSnN$_4$ through FE polarization, presenting a promising candidate for the advancement of sophisticated FE logic storage and spintronic devices. As illustrated in Fig. 4. we propose two operational modes: i) below 200K, when the polarization of AlN bilayer is downward, VSiSnN$_4$ ML is a semiconductor. This is considered as a "0" signal (Fig. 4b). Upon the FE polarization of AlN bilayer is reversed to be upward by applying a voltage, VSiSnN$_4$ ML undergoes a transition into a FM semi-metal state. This transition effectively enhances electron transport through the channel and achieves 100% spin polarization. Subsequently, this is designated as a "1" signal (Fig. 4a). II) The temperature range is between 203K and 284K. The FM-paramagnetic phase transition takes place in VSiSnN$_4$ at 203K, which is attributed to the downward FE polarization of AlN bilayer resulting in a detected "0" signal (Fig. 4b). Conversely, when the FE polarization of AlN bilayer is reversed and directed upwards, VSiSnN$_4$ ML retains its FM properties, indicating a "1" signal (Fig. 4a). The utilization of both operational modes in magnetic memory, as demonstrated here, has been sparsely documented previously. Therefore, our work presents a potential candidate for non-volatile memory with promising implications.

In summary, we propose a vdW heterostructure composed of FM VSiSnN$_4$ ML and FE AlN bilayer. Using systematical first-principles calculations, we demonstrate tunable magnetic anisotropy and switchable semi-metallic in VSiSnN$_4$/AlN heterostructures when the FE polarization of AlN bilayer is reversed from upward to downward. In addition, due to the presence of the upward polarization, the Curie temperature of VSiSnN$_4$ is increased to 284K, which exceeds most of 2D ferromagnets that have been successfully synthesized. Finally, we design a prototype of non-volatile logic memory device based on VSiSnN$_4$/AlN, which provides a potential scheme for realizing and controlling the semi-metallic of 2D ferromagnets and promoting the progress of spintronics.

## SUPPLEMENTARY MATERIAL

See supplementary material for computational methods, introduction of AlN bilayer, schematic drawing of orbital-resolved band structure with SOC of VSiSnN$_4$ monolayer, schematic drawing of VSiSnN$_4$/AlN bilayer, Monte Carlo simulation results of HS-P↑, calculation results of Heisenberg exchange interaction parameters and Binding energy, angle-dependent MCA of VSiSnN4/AlN (Sn-terminal- P↑), details for obtaining orbital resolved MCA.

## ACKNOWLEDGEMENTS

This project is supported by NKRDPC-2022YFA1403300, National Nature Science Foundation of China (No. 12104518, 62104073 and 92165204), GBABRF-2022A1515012643 and GZABRF-202201011118. Yusheng Hou is also supported by the Open Project of Guangdong Provincial Key Laboratory of Magnetoelectric Physics and Devices (No. 2022B1212010008). Density functional theory calculations are performed at Tianhe-II.

# AUTHOR DECLARATIONS

**Conflict of Interest**

The author has no conflict of interest to declare.

**Authors Contributions**

**Kang-Jie Li:** Investigation; Methodology; Writing – original draft. **Ze-Quan Wang:** Methodology; **Zu-Xin Chen:** Conceptualization (equal); Investigation (equal); Funding acquisition (equal); Supervision (equal); Writing – review and editing (equal). **Yusheng Hou:** Conceptualization (equal); Funding acquisition (equal); Investigation (equal); Project administration (equal); Resources (equal); Supervision (equal); Writing – review and editing (equal).

# DATA AVAILABILITY

The data that support the findings of this study are available from the corresponding author upon reasonable request.

# REFERENCES

1. M. M. Vopson, Solid State Communications **152**, 2067 (2012).
2. L. W. Martin and R. Ramesh, Acta Materialia **60**, 2449 (2012).
3. W. Eerenstein, N. D. Mathur and J. F. Scott, Nature **442**, 759 (2006).
4. S.-W. Cheong and M. Mostovoy, Nature Materials **6**, 13 (2007).
5. R. Ramesh and N. A. Spaldin, Nature Materials **6**, 21 (2007).
6. N. Hur, S. Park, P. A. Sharma, J. S. Ahn, S. Guha and S. W. Cheong, Nature **429**, 392 (2004).
7. M. Fiebig, T. Lottermoser, D. Meier and M. Trassin, Nature Reviews Materials **1**, 1 (2016).
8. R. Gupta and R. K. Kotnala, Journal of Materials Science **57**, 12710 (2022).
9. M. Vopsaroiu, J. Blackburn and M. G. Cain, Journal of Physics D-Applied Physics **40**, 5027 (2007).
10. Y. Zhang, Z. Li, C. Deng, J. Ma, Y. Lin and C.-W. Nan, Applied Physics Letters **92**, 152510 (2008).

11. J. Ma, J. Hu, Z. Li and C.-W. Nan, Advanced Materials **23**, 1062 (2011).

12. M. Kumar, S. Shankar, A. Kumar, A. Anshul, M. Jayasimhadri and O. P. Thakur, Journal of Materials Science-Materials in Electronics **31**, 19487 (2020).

13. Z. Chu, M. PourhosseiniAsl and S. Dong, Journal of Physics D-Applied Physics **51**, 243001 (2018).

14. P. Chouhan and R. K. Dwivedi, International Journal of Materials Research **114**, 596 (2023).

15. C.-W. Nan, M. I. Bichurin, S. Dong, D. Viehland and G. Srinivasan, Journal of Applied Physics **103**, 031101 (2008).

16. A. B. Ustinov, A. V. Drozdovskii, A. A. Nikitin, A. A. Semenov, D. A. Bozhko, A. A. Serga, B. Hillebrands, E. Lahderanta and B. A. Kalinikos, Communications Physics **2**, 137 (2019).

17. L. M. Garten, M. L. Staruch, K. Bussmann, J. Wollmershauser and P. Finkel, Acs Applied Materials & Interfaces **14**, 25701 (2022).

18. A. Kumar, A. Goswami, K. Singh, R. McGee, T. Thundat and D. Kaur, Acs Applied Electronic Materials **1**, 2226 (2019).

19. W. Sun, W. Wang, C. Yang, X. Li, H. Li, S. Huang and Z. Cheng, Physical Review B **107**, 184439 (2023).

20. W. Tang, D. Zhao, X. Weng, K. Wu, Z. Yang, C. Kang, Y. Sun, W.-C. Jiang, H. Liang, C. Wang and Y.-J. Zeng, Applied Physics Reviews **10**, 031404 (2023).

21. J. Lu, N. Guo, Y. Duan, S. Wang, Y. Mao, S. Yi, L. Meng, X. Yao and X. Zhang, Physical Chemistry Chemical Physics **25**, 21227 (2023).

22. J. Wang, J. B. Neaton, H. Zheng, V. Nagarajan, S. B. Ogale, B. Liu, D. Viehland, V. Vaithyanathan, D. G. Schlom, U. V. Waghmare, N. A. Spaldin, K. M. Rabe, M. Wuttig and R. Ramesh, Science **299**, 1719 (2003).

23. C. Ederer and N. A. Spaldin, Physical Review B **71**, 060401 (2005).

24. M. Fiebig, Journal of Physics D-Applied Physics **38**, R123 (2005).

25. Y.-H. Chu, L. W. Martin, M. B. Holcomb, M. Gajek, S.-J. Han, Q. He, N. Balke, C.-H. Yang, D. Lee, W. Hu, Q. Zhan, P.-L. Yang, A. Fraile-Rodriguez, A. Scholl, S. X. Wang and R. Ramesh, Nature Materials **7**, 478 (2008).

26. S. Y. Yang, J. Seidel, S. J. Byrnes, P. Shafer, C. H. Yang, M. D. Rossell, P. Yu, Y. H. Chu, J. F.

Scott, J. W. Ager, III, L. W. Martin and R. Ramesh, Nature Nanotechnology **5**, 143 (2010).

27. T. Kimura, T. Goto, H. Shintani, K. Ishizaka, T. Arima and Y. Tokura, Nature **426**, 55 (2003).

28. T. Goto, T. Kimura, G. Lawes, A. P. Ramirez and Y. Tokura, Physical Review Letters **92**, 257201 (2004).

29. M. Kenzelmann, A. B. Harris, S. Jonas, C. Broholm, J. Schefer, S. B. Kim, C. L. Zhang, S. W. Cheong, O. P. Vajk and J. W. Lynn, Physical Review Letters **95**, 087206 (2005).

30. A. Pimenov, A. A. Mukhin, V. Y. Ivanov, V. D. Travkin, A. M. Balbashov and A. Loidl, Nature Physics **2**, 97 (2006).

31. Y. Yamasaki, H. Sagayama, T. Goto, M. Matsuura, K. Hirota, T. Arima and Y. Tokura, Physical Review Letters **98**, 147204 (2007).

32. Y. F. Popov, A. M. Kadomtseva, S. S. Krotov, D. V. Belov, G. P. Vorob'ev, P. N. Makhov and A. K. Zvezdin, Low Temperature Physics **27**, 478 (2001).

33. I. A. Sergienko, C. Sen and E. Dagotto, Physical Review Letters **97**, 227204 (2006).

34. T. Kurumaji, Physical Sciences Reviews **5**, 20190016 (2020).

35. R. E. Newnham, D. P. Skinner and L. E. Cross, Materials Research Bulletin **13**, 325 (1978).

36. B. Huang, G. Clark, E. Navarro-Moratalla, D. R. Klein, R. Cheng, K. L. Seyler, D. Zhong, E. Schmidgall, M. A. McGuire and D. H. Cobden, Nature **546**, 270 (2017).

37. Y. Deng, Y. Yu, Y. Song, J. Zhang, N. Z. Wang, Z. Sun, Y. Yi, Y. Z. Wu, S. Wu and J. Zhu, Nature **563** (7729), 94 (2018).

38. T. Wang, D. Zhang, S. Yang, Z. Lin, Q. Chen, J. Yang, Q. Gong, Z. Chen, Y. Ye and W. Liu, Nature Communications **14**, 5966 (2023).

39. X. Gao, Q. Chen, Q. Qin, L. Li, M. Liu, D. Hao, J. Li, J. Li, Z. Wang and Z. Chen, Nano Research, 1 (2023).

40. F. Liu, L. You, K. L. Seyler, X. Li, P. Yu, J. Lin, X. Wang, J. Zhou, H. Wang and H. He, Nature Communications **7** (1), 1 (2016).

41. Y. Zhou, D. Wu, Y. Zhu, Y. Cho, Q. He, X. Yang, K. Herrera, Z. Chu, Y. Han, M. C. Downer, H. Peng and K. Lai, Nano Letters **17**, 5508 (2017).

42. Q. Wang, L. Cao, S.-J. Liang, W. Wu, G. Wang, C. H. Lee, W. L. Ong, H. Y. Yang, L. K. Ang, S. A. Yang and Y. S. Ang, Npj 2d Materials and Applications **5** (1), 71 (2021).

43. Y.-L. Hong, Z. Liu, L. Wang, T. Zhou, W. Ma, C. Xu, S. Feng, L. Chen, M.-L. Chen, D.-M. Sun, X.-Q. Chen, H.-M. Cheng and W. Ren, Science **369**, 670 (2020).

44. L. Wang, Y. Shi, M. Liu, A. Zhang, Y.-L. Hong, R. Li, Q. Gao, M. Chen, W. Ren, H.-M. Cheng, Y. Li and X.-Q. Chen, Nature Communications **12**, 2361 (2021).

45. S.-D. Guo, Y.-T. Zhu, W.-Q. Mu and W.-C. Ren, Europhysics Letters **132**, 57002 (2020).

46. Q. Cui, Y. Zhu, J. Liang, P. Cui and H. Yang, Physical Review B **103** (8), 085421 (2021).

47. S. Han, S. Lee, D. Ko, X. Zhang, J. Kim, C. A. Ross and D. H. Kim, Advanced Functional Materials, 2306909 (2023).

48. Y. Wang, X. Xu, W. Ji, S. Li, Y. Li and X. Zhao, Npj Computational Materials **9** (1) (2023).

49. A. Konishi, T. Ogawa, C. A. J. Fisher, A. Kuwabara, T. Shimizu, S. Yasui, M. Itoh and H. Moriwake, Applied Physics Letters **109**, 150 (2016).

50. Z. Wang, G. Wang, X. Liu, S. Wang, T. Wang, S. Zhang, J. Yu, G. Zhao and L. Zhang, Journal of Materials Chemistry C **9**, 17201 (2021).

51. L. Li and M. Wu, Acs Nano **11**, 6382 (2017).

52. P. Li, X. Yang, Q.-S. Jiang, Y.-T. Wu and W. Xun, Physical Review Materials **7**, 064002 (2023).

53. D. Dey, A. Ray and L. Yu, Physical Review Materials **6**, L061002 (2022).

54. Q. Cui, J. Liang, Z. Shao, P. Cui and H. Yang, Physical Review B **102**, 094425 (2020).

55. F. Xue, Y. Hou, Z. Wang and R. Wu, Physical Review B **100**, 224429 (2019).

56. S. Zhang, R. Xu, W. Duan and X. Zou, Advanced Functional Materials **29**, 1808380 (2019).

57. L. Ju, Y. Dai, W. Wei, M. Li and B. Huang, Applied Surface Science **434**, 365 (2018).

58. Wang, Wu and Freeman, Physical review. B, Condensed matter **47**, 14932 (1993).

59. Y.-Q. Li, X. Zhang, X. Shang, Q.-W. He, D.-S. Tang, X.-C. Wang and C.-G. Duan, Nano Letters **23**, 10013 (2023).